\documentclass[doublecol]{epl2} 
\usepackage[utf8x]{inputenc}\SetUnicodeOption{mathletters} \SetUnicodeOption{autogenerated}

\title{Spin-orbit interaction from low-symmetry localized defects in semiconductors}

\author{Oleg Chalaev\inst{1} \and G. Vignale\inst{1}  \and Michael E. Flatt\'e\inst{2}}
\institute{%
  \inst{1} Department of Physics and Astronomy, University of Missouri, Columbia, MO 65211, USA\\
  \inst{2} Optical Science and Technology Center and Department of Physics and Astronomy, University of Iowa, Iowa City, Iowa, 52242, USA}

\shortauthor{Chalaev, Vignale and Flatt\'e}


\abstract{The presence of low-symmetry impurities or defect complexes in the zinc-blende direct-gap semiconductors (e.g. interstitials, DX-centers) results in a novel spin-orbit term in
the effective Hamiltonian for the conduction band.  The new extrinsic spin-orbit interaction is proportional to the matrix element of the defect potential between the conduction and the
valence bands.  Because this interaction arises already in the first order of the expansion of the effective Hamiltonian in powers of $U_{\mathrm{ext}}/E_g\ll 1$ (where~$U_{\mathrm{ext}}$
is the pseudopotential of an interstitial atom, and
$E_g$ is the band gap), its contribution to the spin relaxation rate may exceed that of the previously studied extrinsic contributions, even for moderate concentrations of impurities.}

\usepackage{graphicx,amssymb,amsmath,amsbsy,subfigure,dsfont,ifthen,ifpdf,color,subscript,marvosym,bbm,array}
\newcommand{\Sp}{\mathop{\mathrm{Tr}}}

\newcommand{\ud}{\mathrm d}

\newcommand{\elib}[1]{\ifpdf \href{file://#1}{\includegraphics[height=.5cm]{images/dvd}}\else{\includegraphics[height=2ex]{images/dvd}}\fi}

\newcommand{\insfig}[2]{\begin{minipage}[c]{#1}\includegraphics[width=#1]{#2}\end{minipage}}

\DeclareUnicodeCharacter{9786}{\Smiley}
\DeclareUnicodeCharacter{9785}{\frownie}
\DeclareUnicodeCharacter{411}{\lambdaslash}
\DeclareUnicodeCharacter{9001}{\left\langle}
\DeclareUnicodeCharacter{9002}{\right\rangle}
\DeclareUnicodeCharacter{8658}{\Longrightarrow}
\DeclareUnicodeCharacter{183}{\cdot}
\DeclareUnicodeCharacter{8469}{\mathbb N}

\begin{document}
\date{December 31, 2011}

\maketitle
\section{Introduction}
The spin-orbit interaction (SOI) is a remarkable example of an ``emergent interaction''.  For electrons in vacuum, the SOI
emerges from the projection of the Dirac equation on positive energy states.  It is intrinsically weak, since its ``coupling constant'' contains the electron rest energy $mc^2$ in the
denominator, and thus requires strong nuclear potentials to become truly effective.  For electrons in the conduction band of semiconductors, the situation is different.  The spin and
orbital parts of the Bloch wave functions for these electrons are already entangled by the spin-orbit interaction arising from the strong periodic potential in the atomic cores.
Therefore, the potential from a symmetry-breaking defect leads to a SOI that is orders of magnitude larger~\cite{Blount1962,Winkler,Engel} than the SOI one would expect in vacuum for the same strength of
the potential.  This enhanced SOI plays a crucial role in recently proposed schemes for the electric manipulation of spins in semiconductors~\cite{Awschalom2002,Awschalom2007}.  So far, however, the attention has been focused on defects that break the translational symmetry of the lattice (e.g. substitutional defects, vacancies) leaving the point group intact. 
But the breaking of translational symmetry, as will be shown below, adds the defect potential $U_{\mathrm{ext}}$ to the
{\it diagonal} elements of the Hamiltonian~\cite{Winkler,PikusTitkov}, whose conduction and valence states are already split by the band gap~$E_g\gg U_{\mathrm{ext}}$,
and leads only to relatively small corrections (the so-called ``central cell corrections''~\cite{YuCardona}) to energy spectra, effective electron mass, etc.
Contrary to that, the point-group symmetry reduction results in a correction to the {\it off-diagonal} matrix elements (between the conduction and the valence band) of the Hamiltonian,
which are small to begin with; consequently the corresponding corrections are \emph{not} expected to be relatively small, especially
in a wide-gap
semiconductor where one might expect the conduction-band SOI to be more sensitive
to a lowering of the point group symmetry of an atomic cell than to the breaking of translational symmetry.  

In this Letter we demonstrate that  lowering  the point group symmetry leads to a \emph{novel type of spin-orbit interaction} [see Eq.~\eqref{novelSOI} below] in direct-gap
semiconductors, and we explore some implications for the practically important calculation of the spin relaxation rate~\cite{Elliott1954,Yafet1963,PikusTitkov,Lau2001,PhysRevB.64.201301,Wu201061,ActaFabian}.
The reduction in point group symmetry provides the leading-order contribution to spin relaxation whenever the conventional
D'yakonov-Perel' (DP) mechanism~\cite{PikusTitkov} due to the spin-split band structure is absent (this happens, for example, in semiconductor quantum wells grown along certain  directions~\cite{Dyakonov1986}).  We will demonstrate this point by
explicitly comparing the spin relaxation rate produced by interstitial defects with the spin relaxation rate produced by substitutional defects with a  comparable elastic relaxation time~$\tau $.
Furthermore, even though the DP mechanism is frequently thought to be the dominant mechanism for spin relaxation in III-V semiconductors (both in the bulk and in quantum
wells) \cite{Lau2001,Wu201061,ActaFabian}, we find that in some cases the dominance of DP is not so clear-cut: our mechanism can compete and even dominate for strong enough disorder, so long as the disorder does not invalidate the diffusive transport regime.  

A direct-gap semiconductor subjected to an extrinsic field~$U_{\mathrm{ext}}(\vec r\,)$ is described by the Hamiltonian
\begin{equation}\label{hamiltonian}
{\hat H}=\frac{{\hat p}^2}{2m}+U_{\mathrm{at}}(\hat{\vec r}\,)+U_{\mathrm{ext}}(\hat{\vec r}\,),
\end{equation}
where $\hat{\vec p}=-i\hbar {\vec{\nabla }}\!_r$ is the momentum operator, $U_{\mathrm{at}}$ is the periodic crystal potential and $U_{\mathrm{ext}}$ is an ``extrinsic'' potential from impurities, defects, and/or other sources. 

In the standard $\mbox{$\vec k\!\cdot \!\hat{\vec p}$}$ approach, which we adopt here, the Hamiltonian is represented as a matrix on the basis of wave functions $\psi _{n\vec k}(\vec r\,) \equiv e^{i\vec k\cdot \vec r}u_{n0}(\vec r\,)$, where $u_{n0}(\vec r\,)$ are the periodic parts of the Bloch wave functions of the periodic crystal at $\vec k=0$.  Notice that $\psi _{n\vec k}$ are not  Bloch wave functions when $\vec k \ne 0$: for this reason, even in the absence of the external potential, the Hamiltonian has off-diagonal matrix elements of the form $\vec k\cdot \vec p_{nn'}\delta _{\vec k \vec k'}/m$ between $\psi _{n\vec k}$ and $\psi _{n'\vec k'}$, with $n \ne  n'$.  Here  $\vec p_{nn'}$ is the matrix element of $\hat{\vec p}$ between $u_{n0}(\vec r\,)$ and $u_{n'0}(\vec r\,)$. 
The field~$U_{\mathrm{ext}}(\vec r\,)$ lowers the otherwise perfect symmetry of the crystal in two ways:
\begin{enumerate}
\item It breaks the translational symmetry by creating diagonal (intra-band) matrix elements of 
 the form $U_{\mathrm{ext}}(\vec k-\vec k')\delta _{nn'}$, where $U_{\mathrm{ext}}(\vec k)$ is the Fourier transform of  $U_{\mathrm{ext}}(\vec r\,)$; 
\item It may also  lower the local point group symmetry,  producing additional off-diagonal (inter-band) matrix elements with $n \ne  n'$ and $\vec k = \vec k'$.  
\end{enumerate}
An  interstitial atom or a point defect such as a DX-center\cite{YuCardona} or an EL2-complex~\cite{Pantelides-dc}
affects both symmetries: the translational one and the point-group one.  The traditional approach to the generation of spin-orbit interaction focuses on the intra-band part of $U_{\mathrm{ext}}$, which arises from the breaking of translational symmetry.  Here we concentrate on the inter-band part of $U_{\mathrm{ext}}$, which arises from the lowering of point-group symmetry.

Let us consider a situation in which the extrinsic field~$U_{\mathrm{ext}}$ is produced by short-range interstitial defects, e.g., DX-centers, which are known to appear spontaneously in
Al\textsubscript xGa\textsubscript{1-x}As for $0.22<x<0.5$~\cite{Queisser14081998}.   We  assume that the field induced by every single defect is negligibly small outside of the atomic
cell in which the defect resides.

This assumption makes it convenient to separate
the coordinate variable~$\vec r$  (determining an arbitrary point inside the crystal)
into a ``macroscopic'' ($\vec\rho$\,) and a ``microscopic''($\vec{\bar r}$\,) part: $\vec\rho$ is a discrete variable pointing to the origin of each unit cell, while   $\vec{\bar r}$ varies within the cell.  Any point inside the crystal is thus uniquely identified by the pair of 
variables~$(\vec\rho,\vec{\bar r}\,)$ so that $\vec r=\vec\rho+\vec{\bar r}$.
With this notation
our external  potential  can now be factorized into the  product of a macroscopic part, $U(\vec\rho\,)$,  and a (dimensionless) microscopic part, $V(\vec{\bar r}\,)$:
\begin{equation}\label{disModel}
U_{\mathrm{ext}}(\vec\rho,\vec{\bar r}\,)=U(\vec\rho\,)V(\vec{\bar r}\,),\quad\int _vV(\vec{\bar r}\,)\ud^3r=|v|,
\end{equation}
where $v$ denotes the unit cell centered at the origin, and~$|v|$ is its volume.
Let us also assume that the defects are placed into a randomly chosen set of unit
cells, centered at positions $\vec\rho_j$ with $j=1\ldots{}n$. Then we can write 
\begin{equation}\label{modelBesp}
    U(\vec\rho\,)\!=\!U^{(0)}\!\sum _{j=1}^{n}\chi (\vec\rho-\!{\vec\rho}_{j}),\ \  \chi (\vec\rho\,)\!=\!\begin{cases}1/|v|, &\vec\rho\in v,\\0, &\vec\rho\notin v,\end{cases}
\end{equation}
where $n$ is the total number of  defects, and~$U^{(0)}$ is the amplitude of the potential.

In the macroscopic limit (which corresponds to the electron gas model)~$|v|\rightarrow 0$, so we have  $\chi (\vec\rho-{\vec\rho}_{j})\rightarrow \delta (\vec\rho-{\vec\rho}_{j})$
in~\eqref{modelBesp} and the disorder potential is described by the standard Lifshitz disorder model~\cite{montambauxBook}
\begin{equation}\label{Lifshitz}
\overline{U(\vec\rho\,)U(\vec\rho\,')}=\frac{\hbar }{2\pi \nu _E\tau _E}\delta (\vec\rho-\vec\rho\,')\,,
\end{equation}
where~$\nu _E$ is the density of states at the considered electron energy, $\tau _E$ is the elastic momentum relaxation time at the same energy, and the overline stands for the averaging over different disorder configurations.

In what follows we will be concerned with (direct-gap) zinc-blende semiconductors, for which the two-fold degenerate conduction band is described by the $\Gamma _6$-irreducible representation
(irrep) of the double cubic group, while the six-fold degenerate valence band is a mixture of two subbands described by irreps~$\Gamma _8$ and~$\Gamma _7$.

In this Letter we use the $8\times 8$ $\vec k\cdot \vec p$ model, where the initial Hamiltonian~\eqref{hamiltonian} is
reduced to the $8\times 8$-matrix ${\hat H}_{\vec\rho}$ whose elements are operators acting on $\vec\rho$-dependent functions (where $\vec\rho$ is a long-range variable defined above).
Let us divide the $8\times 8$ matrix ${\hat H}_{\vec\rho}$ into four blocks as follows:\cite{approxes}
\begin{equation}\label{kpHam}  \begin{split}
    {\hat H}_{\vec\rho}=\begin{pmatrix}  {\hat{\bar H}}_c&{\hat{\bar H}}_{cv}\\{\hat{\bar H}}_{cv}^\dag{}&{\hat{\bar H}}_v\end{pmatrix},\quad
    {\hat{\bar H}}_v=\begin{pmatrix}  {\hat{\bar H}}_{v8}&0\\0&{\hat{\bar H}}_{v7}\end{pmatrix},\\
{\hat{\bar H}}_c\approx {\hat{\bar H}}_{v8}+E_g,\quad {\hat{\bar H}}_{v8}\approx {\hat{\bar H}}_{v7}+\Delta ,
  \end{split}\end{equation}
where $\Delta $ is the spin-orbit-induced splitting between the $\Gamma _8$ and $\Gamma _7$ valence bands; the blocks ${\hat{\bar H}}_c$ and~${\hat{\bar H}}_{v7}$ are two-dimensional, and the block ${\hat{\bar H}}_{v8}$ is four-dimensional.
Using the well-known mathematical formula for the inverse of a block matrix, we express the $8\times 8$ Green's function (GF)
${\hat G}^E\equiv [E-\hat H]^{-1}$ in terms of the matrix blocks defined in~\eqref{kpHam}.

When holes in the valence band are absent, all transport properties are controlled by electrons in the conduction band.  Under these conditions the relevant information from 
the  full GF ${\hat G}^E$ is contained entirely in the upper-left $2\times 2$ CB-block ${\hat G}^E_c$ of ${\hat G}^E$:
\begin{equation}\label{CBGF}
[{\hat G}^E_c]^{-1}=E-{\hat{\bar H}}_c+{\hat{\bar H}}_{cv}\left[E-{\hat{\bar H}}_v\right]^{-1}{\hat{\bar H}}_{cv}^\dag{}.
\end{equation}
The conduction-band GF ${\hat G}^E_c$ corresponds to the effective conduction-band Hamiltonian according to the usual rule
\begin{equation}\label{condBandHam}
{\hat G}^E_c=[E-\hat{\tilde H}\vphantom{H}^E]^{-1},\quad \hat{\tilde H}\vphantom{H}^E=E-({\hat{\tilde G}}\vphantom{G}^E)^{-1}.
\end{equation}
The spin-dependent part of the Hamiltonian $\hat{\tilde H}\vphantom{H}^E$
is entirely contained within the last term in~\eqref{CBGF}.
In the absence of  spin-orbit splitting of the valence bands  (i.e.,~for $\Delta =0$)  $\hat{\tilde H}\vphantom{H}^E$ would be spin-independent since
the following sum (over \emph{all considered}\cite{allConsidered} states) is  spin-independent:
\begin{equation}\label{sumRule}
\sum _{l,n\in \Gamma _{7,8}}{\hat{\bar H}}_{cv}|l\rangle\langle l|\left[E-{\hat{\bar H}}_v\right]^{-1}|n\rangle\langle n|{\hat{\bar H}}_{cv}^\dag{},
\end{equation}
where the sum is taken over \emph{envelope} states in the valence bands.
[We note that due to the diagonal form of~${\hat{\bar H}}_v$ in~\eqref{kpHam} only terms with $l=n$ survive in~\eqref{sumRule}.]
Due to the sum rule~\eqref{sumRule} it is possible to express the spin-dependent part of~$\hat{\tilde H}\vphantom{H}^E$ as a sum over
either $\Gamma _7$ or $\Gamma _8$ states. We prefer to sum over $\Gamma _7$ states since this results in simpler expressions; then the spin-dependent part of~$\hat{\tilde H}\vphantom{H}^E$ is
entirely contained in
  \begin{equation}\label{tosdp}    \begin{split}
  &\sum _{n\in \Gamma _7}\left[\frac{\hat{\vec k}\hat{\vec p}}m+U_{\mathrm{ext}}(\vec r\,)\right]\left|n\right\rangle \\
  &\left\langle n|{\hat{\mathbbm G}}^{E-E_g-\Delta }-{\hat{\mathbbm G}}^{E-E_g}|n\right\rangle  \left\langle n\right|\left[\frac{\hat{\vec k}\hat{\vec p}}m+U_{\mathrm{ext}}(\vec r\,)\right],
    \end{split}  \end{equation}
where~${\hat{\mathbbm G}}^E$ is a scalar GF:
\begin{equation}
{\hat{\mathbbm G}}^E\equiv \left[E-\frac{{\hat p}^2}{2m}-U_{\mathrm{ext}}(\vec r\,)\right]^{-1}\!\!\!\!\!\!,
\end{equation}
and the energy zero corresponds to the bottom of the conduction band.
Different spin-orbit contributions can be obtained by expanding ${\hat{\mathbbm G}}^{E-E_g-\Delta }-{\hat{\mathbbm G}}^{E-E_g}$ in~\eqref{tosdp}
in powers of~$\left\langle \Gamma _7\left|U_{\mathrm{ext}}\right|\Gamma _7\right\rangle /E_g\ll 1$ and~$\left\langle \Gamma _7\left|U_{\mathrm{ext}}\right|\Gamma _7\right\rangle /(E_g+\Delta )\ll 1$,
where $\vec\rho$-dependent
$\left\langle \Gamma _7\left|U_{\mathrm{ext}}\right|\Gamma _7\right\rangle $ is a $2\times 2$ matrix obtained by projecting the field~$U_{\mathrm{ext}}$ onto the $\Gamma _7$-states.
Thus in our calculation the quantity $\left\langle \Gamma _7\left|U_{\mathrm{ext}}\right|\Gamma _7\right\rangle $
must be understood as the pseudopotential associated with the
point-group-symmetry-breaking defect; the importance of these symmetry-breaking potentials extends to regimes of larger
$U_{\mathrm{ext}}(\vec\rho\,)$, for which the higher-order terms in the expansion of $G_c^E$
can sometimes be analytically summed~\cite{Pantelides-dc}.

\section{The novel spin-orbit interaction}
The zeroth order of the  expansion corresponds to the approximations $|E-E_{\mathrm F}|\ll E_g$ (where $E_{\mathrm F}$ is the Fermi energy) and~$U_{\mathrm{ext}}\ll E_g$
in~$\hat{\mathbbm G}$, or, in other words,
\begin{equation}\label{zoap}
{\hat{\mathbbm G}}^{E-E_g-\Delta }-{\hat{\mathbbm G}}^{E-E_g}\approx \frac1{E_g}-\frac1{E_g+\Delta }.
\end{equation}
Substituting~\eqref{zoap} in~\eqref{tosdp} and disregarding all spin-independent terms one obtains
\begin{equation}\label{zerothOrder}
{\hat H}_{\mathrm{SOI}}=\frac13\left[\frac1{E_g}-\frac1{E_g+\Delta }\right]{\hat H}_{cv7}{\hat H}_{cv7}^\dag{},
\end{equation}
where ${\hat H}_{c,v7}$ is the {rightmost} $2\times 2$ part of ${\hat{\bar H}}_{cv}$ defined in~\eqref{kpHam}  (in other words,
${\hat H}_{c,v7}$ is the Hamiltonian matrix taken between the conduction band states $\Gamma _6$ and the valence band states $\Gamma _7$).
Making use of Eq.~(\ref{disModel}) for the
external potential these matrix elements work out to be
\begin{equation}\label{Hcv}
  \begin{split}
    [{\hat H}_{c,v7}]_{\vec k,\vec k'}\equiv \langle \Gamma _6|e^{-i\vec k\cdot \vec r}\hat H e^{i\vec k'\cdot \vec r}|\Gamma _7\rangle=\\
-\vec{\sigma } \cdot \left[\frac{iP{\vec k}}{\hbar }\delta _{\vec k,\vec k'}+U_{\vec k-\vec k'}{\vec V}\right],
  \end{split}
\end{equation}
where $P\equiv P_7=i\hbar \langle S|{\hat p}_x|X_7\rangle/m$ and  $\vec V\equiv \left\langle  S\left|V\right|{\vec R}\right\rangle $.   Here  $|S\rangle$ denotes the orbital part of the $\Gamma _6$ state,
and the components of the (spin-independent) vector function $|{\vec R}\rangle\equiv \left(|X_7\rangle,|Y_7\rangle,|Z_7\rangle\right)$
participate in the basis formation for the irrep~$\Gamma _7$.
Lastly, the vector~$\vec{\sigma }\equiv (\sigma _1,\sigma _2,\sigma _3)$ denotes the set of three  Pauli matrices.
From Eqs.~\eqref{zerothOrder} and \eqref{Hcv} we extract the spin-dependent part of~$\hat{\tilde H}$:
\begin{eqnarray}\label{novelSOI}
{\hat H}_{\mathrm{SOI}}=
-i\frac{l_1}{\hbar }\ \hat{\vec s}\cdot \left[\vec V\times \left({\vec{\nabla }}_{\vec\rho} U(\vec\rho)+U(\vec\rho){\vec{\nabla }}_{\vec\rho}\right)\right]\,,
\end{eqnarray}
where $\hat{\vec s}\equiv \hbar \vec{\sigma }/2$ is the spin operator and 
\begin{equation}\label{nsLength}
l_1\equiv \frac{2P}3\left(\frac1{E_g}-\frac1{E_g+\Delta }\right).
\end{equation}
Notice that in order to arrive at  Eq.~\eqref{novelSOI} we replaced the wave vector $\vec k$ of Eq.~(\ref{Hcv}) by the operator ${\vec{\nabla }}_{\vec\rho}$.
This is legitimate because the basis functions of our representation depend on $\vec k$ only through the plane wave factor $e^{i\vec k\cdot \vec r}$ which,
in the macroscopic limit, becomes~$e^{i\vec k\cdot \vec\rho}$, since $\vec k \cdot \vec {\bar r}\ll 1$. Thus, on this basis, the operator ${\vec{\nabla }}_{\vec\rho}$ 
is equivalent to multiplication by~$i\vec k$.  

\section{The Rashba spin-orbit interaction}
The first order of the  expansion of~${\hat{\mathbbm G}}^{E-E_g-\Delta }-{\hat{\mathbbm G}}^{E-E_g}$ in~\eqref{tosdp} in powers of~$U_{\mathrm{ext}}$ corresponds to the approximation
\begin{equation}\label{foeiU}
  \begin{split}
{\hat{\mathbbm G}}^{E-E_g-\Delta }\approx &\frac1{E_g-\Delta }\left\langle \Gamma _7\left|U_{\mathrm{ext}}\right|\Gamma _7\right\rangle \frac1{E_g-\Delta },\\
{\hat{\mathbbm G}}^{E-E_g}&\approx \frac1{E_g}\left\langle \Gamma _7\left|U_{\mathrm{ext}}\right|\Gamma _7\right\rangle \frac1{E_g},
  \end{split}
\end{equation}
where~$\left\langle \Gamma _7\left|U_{\mathrm{ext}}\right|\Gamma _7\right\rangle $ is a $2\times 2$ matrix obtained by projecting the field~$U_{\mathrm{ext}}$ onto the $\Gamma _7$-states.
Similarly to how it has been done in the previous section, but using~\eqref{foeiU} instead of~\eqref{zoap}, one reproduces the Rashba spin-orbit term:
\begin{equation}\label{RashbaResult}  \begin{split}
    {\hat H}_{\mathrm R}\!=\!\frac{l^2_2}{2\hbar ^2}(\hat{\vec k}\vec{\sigma })U(\vec\rho\,)(\hat{\vec k}\vec{\sigma })
\!=\!\frac{l^2_2}{\hbar ^2}\hat{\vec s}\cdot \!\left[(\vec{\nabla }_{\vec\rho} U(\vec\rho))\times \hat{\vec k}\right],\\
    l^2_2=\frac{2P^2}3\left[E_g^{-2}-(E_g+\Delta )^{-2}\right],
  \end{split}\end{equation}
The Rashba result~\eqref{RashbaResult} is reproduced in case when $U_{\mathrm{ext}}(\vec r)$ is a smooth function
so that in~\eqref{disModel} $V(\vec{\bar r}\,)\equiv 1$ and~$U_{\mathrm{ext}}(\vec r)\equiv U_{\mathrm{ext}}(\vec\rho,\vec{\bar r}\,)=U(\vec\rho\,)$.
In a more complicated situation, when $U_{\mathrm{ext}}$ may significantly change on the scale of a single atomic cell, central cell corrections arise, which are, however,
small compared to both~\eqref{novelSOI} and~\eqref{RashbaResult}.

\section{Discussion}
The values of the SOI lengths~$l_1$ and~$l_2$ for several direct-gap A$_3$B$_5$ semiconductors are listed in Table~\ref{estab}.
The spin-orbit interaction in~\eqref{novelSOI} is different from the previously studied
extrinsic SOI~\eqref{RashbaResult} in that
 (i) its amplitude contains only one power of $E_g$ in the denominator and
(ii) it  occurs due to defects that violate the \emph{point-group symmetry}  of an atomic cell. 
Point (ii)  is a consequence of the fact that the matrix elements $\vec V$, defined in Eq.~(\ref{Hcv}) are essentially null if the ``form factor'' $V(\bar r)$ of the impurity potential
has the symmetry of the point group of the lattice, as is the case for a substitutional impurity in a zincblende semiconductor.   Thus, the spin-orbit interaction associated with substitutional impurities is of the conventional
Rashba-type~\eqref{RashbaResult}, while the spin-orbit interaction associated with symmetry-lowering defects has an additional inter-band term.

In the rest of the article we see how  the new spin-orbit term~\eqref{novelSOI} contributes to the spin relaxation.

\section{Electron spin relaxation}
\begin{figure}  \centering
\subfigure[]{\includegraphics[width=15ex]{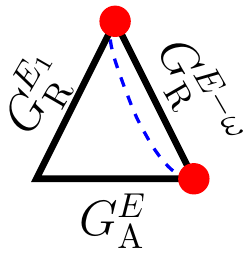}}\qquad\subfigure[]{\includegraphics[width=15ex]{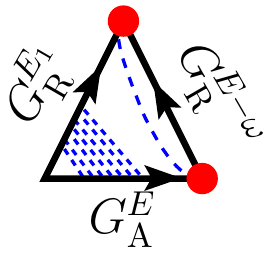}}%
\caption{Two diagrams for the kinetic term in Eq.~\eqref{gkeII} in the leading order of $\hbar /(E\tau )\ll 1 $ expansion (zero loop)~\cite{bigCond}.
The solid lines denote Green's functions, the dashed lines represent averaging over the different disorder realizations; the ladder composed by the dashed lines stands for the
cooperon~\cite{montambauxBook}. The solid circles represent the new SOI interaction of Eq.~\eqref{novelSOI}.}\label{fig:diags}
\end{figure}

We assume that the electrons are initially spin-polarized in the $z$-direction, with  spin distribution function $f_E=f^{(0)}_E\sigma _0+f^{(3)}_E\sigma _3$.
Following the approach outlined in Ref.~\cite{ahiezerEng} we obtain the dynamical equation for~$f^{(3)}_E$: ${\dot f}^{(3)}_E=-f^{(3)}_E/\tau _{\mathrm{SOI}}(E)$,
where the relaxation rate $\tau ^{-1}_{\mathrm{SOI}}(E)$ is given by
\begin{equation}\label{gkeII}  \begin{split}
\tau ^{-1}_{\mathrm{SOI}}&(E_1)=\frac i{\hbar ^2\nu _{E_1}}\int _{-\infty }^\infty \frac{\ud E\ud \omega }{(2\pi )^3}\int _{-\infty }^0\ud t'e^{-i\omega t'/\hbar }\\
    \times \Sp&\Big[\left({\hat G}^{E_1}_{\mathrm R}-{\hat G}^{E_1}_{\mathrm A}\right)^T
\left\{{\hat H}_{\mathrm{SOI}},{\hat G}^{E-\omega }_{\mathrm R}{\hat H}_{\mathrm{SOI}}{\hat G}^E_{\mathrm A}\right\}\Big],
  \end{split}\end{equation}
where ${\hat G}^E_{\mathrm{R/A}}=\left[E+\hbar ^2\nabla ^2_{\vec\rho}/2m-U(\vec\rho\,)\pm i0\right]^{-1}$
are the Green's functions operators for the electron gas in the presence of disorder~\eqref{disModel}.
We calculate the relaxation rate~\eqref{gkeII} using the disorder averaging technique~\cite{montambauxBook} together with the loop expansion~\cite{bigCond}, i.e., the expansion in powers of $\hbar /E\tau _E\ll 1$.
In a degenerate electron gas  $f^{(3)}_E\ne 0$ only in the close vicinity of the Fermi energy~$E_{\mathrm F}$, so that only electrons with energies~$E$ close to~$E_{\mathrm F}$ participate in the relaxation.
In this case, the spin relaxation is exponential with the relaxation rate $\tau ^{-1}_{\mathrm{SOI}}(E_{\mathrm F})$:
\begin{equation}\label{taus}
  {\dot s}_z=-\hbar \int _0^\infty f^{(3)}_E\nu _E\tau ^{-1}_{\mathrm{SOI}}(E)\ud E=-s_z\tau ^{-1}_{\mathrm{SOI}}(E_{\mathrm F}).
\end{equation}
Fig.~\ref{fig:diags} shows the relevant diagrams in the zeroth order of the loop expansion~\cite{bigCond}.%
Both diagrams have a common part: a Green's function with two spin-orbit vertices connected by a disorder-averaging (dashed) line.
First we calculate the integrals over $\omega $ and $t$ from this common part:
\begin{equation}\label{sintsa}
i\Sp~\Im m \int ^0_{-\infty }\!\!\!\!\ud t\int _{-\infty }^\infty \frac{\ud \omega }{2\pi }e^{-i\omega t/\hbar }\insfig{8ex}{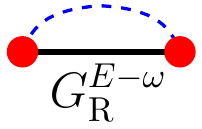}=-\frac{i\hbar ^3x^2_{\mathrm{SOI}}}{2\tau _E},
\end{equation}
where we denoted~$x_{\mathrm{SOI}}\equiv V_zpl_1/\hbar \ll 1$, and~$p=\sqrt{2mE}$.
After the integration in~\eqref{sintsa}, the two triangle diagrams in Fig.~\ref{fig:diags} turn into two bubbles; from~\eqref{gkeII} and~\eqref{sintsa} we obtain
\begin{equation}\label{gkeIIIa}
\tau ^{-1}_{\mathrm{SOI}}(E_1)=\frac{x^2_{\mathrm{SOI}}}{2\tau _{E_1}\nu _{E_1}}\int _{-\infty }^\infty \frac{\ud E}{(2\pi )^2}\left[\insfig{6ex}{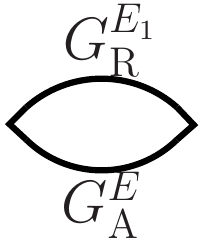}+\insfig{6ex}{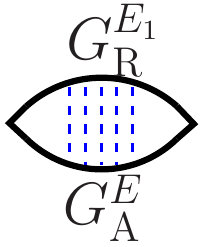}\right],
\end{equation}
so that the spin relaxation is:
\begin{equation}\label{tauBetaA}
  \begin{split}
\text{in 2D}\quad    \tau ^{-1}_{\mathrm{SOI}}(E)&=\frac{4mEl_1^2}{\hbar ^2\tau }V^2_z,\\
\text{in 3D}\quad    \tau ^{-1}_{\mathrm{SOI}}(E)&=\frac{4mEl_1^2}{\hbar ^2\tau }\frac{V^2_x+V^2_y+2V^2_z}{3},
  \end{split}
\end{equation}
where~$\tau $ is the elastic relaxation time.  Here by ``2D'' we mean the two-dimensional electron gas in a quantum well with only one subband occupied by doped electrons, while ``3D" refers to the usual bulk case.

For quantum wells grown along certain particular directions (eg., the [110] direction in GaAs) intrinsic spin precession is absent for spins oriented along a specific direction, and the
extrinsic relaxation mechanism dominates.  The ensuing spin relaxation rate $2V^2_zp^2_{\mathrm F}l_1^2\tau ^{-1}/\hbar ^2$ is proportional to the elastic scattering rate~$\tau ^{-1}\equiv \tau ^{-1}_{E_{\mathrm F}}$.
It is instructive to compare the spin relaxation time arising from the novel SOI~\eqref{novelSOI} with what would be obtained from the conventional approach~\cite{Elliott1954,Yafet1963}, in which the spin-orbit interaction is given by
\begin{equation}
\hat H_{\mathrm{SOI},2} =il_2^2\ \hat{\vec s}\cdot \left[{\vec{\nabla }}_{\vec\rho} U(\vec\rho)\times {\vec{\nabla }}_{\vec\rho}\right]/\hbar ,
\end{equation}
with $l^2_2=2P^2\left[E_g^{-2}-(E_g+\Delta )^{-2}\right]/3$ (see Table I).
Both mechanisms give a spin relaxation rate proportional to $\tau ^{-1} $, so the ratio of the two contributions is independent of $\tau $ and given by
\begin{equation}
\tau _{\mathrm{SOI}}/\tau _{\mathrm{SOI},2} %
 \sim l_2^4/(2a^2l_1^2) \sim 8\times 10^{-3}
\end{equation}
for a GaAs quantum well of width $a=100$\AA{}.
 We see that in 2D the relaxation rate due to the novel SOI is two orders of magnitude larger than the ``conventional'' one.
In the 3D case the ratio $\tau _{\mathrm{SOI}}/\tau _{\mathrm{SOI},2}$ becomes energy-dependent:
\begin{equation}
\frac{\tau _{\mathrm{SOI}}}{\tau _{\mathrm{SOI},2}}\sim \frac{mEl_2^4}{\hbar ^2l_1^2}\sim 0.04 \text{ at $E$=0.03eV}.
\end{equation}
Finally, let us  compare our spin relaxation time with the well-known DP relaxation time, which is caused by the intrinsic spin splitting of the conduction bands due to bulk inversion asymmetry (BIA).  Our calculation in this case gives
\begin{equation}\label{BIArates}
  \begin{split}
\text{in 2D}\quad    \tau ^{-1}_{\mathrm{BIA}}(E)&=\tau _E\frac{32\pi ^4}{a^4}mEb^2,\\
\text{in 3D}\quad    \tau ^{-1}_{\mathrm{BIA}}(E)&=\tau _E\frac{64}{105}\frac{b^2m^3E^3}{\hbar ^8},
  \end{split}\end{equation}
where the parameter~$b\equiv b_{41}^{6c6c}$ is defined in Eq.~(6.2a) of Ref.~\cite{Winkler} and its value is listed in Table I.
Comparing Eqs.~\eqref{tauBetaA} and~\eqref{BIArates} we see that the ratio of the  BIA and SOI relaxation rates is given by
\begin{equation}\label{vle}
\frac{\tau ^{-1}_{\mathrm{BIA}}(E)}{\tau ^{-1}_{\mathrm{SOI}}(E)}= (CE\tau _E/\hbar )^2\,,
\end{equation}
where the constant $C=(\sqrt{1536}/{105})(bm/{l_1\hbar ^2})$ is listed in the last line of Table I for a few semiconductors. 
A short elastic scattering time favors our extrinsic mechanism; on the other hand $\tau _E$ cannot be made too small in our calculation since our diagrammatic approach is valid only for weak disorder, $E\tau _E>\hbar $.   The value of the constant $C$  is reported in Table I for several bulk zinc-blende semiconductors.  Since $C<1$ in all cases, we can conclude that the novel SOI~\eqref{novelSOI} will dominate the spin relaxation rate in a window of parameters such that the condition $1<E\tau _E/\hbar <C^{-1}$ is satisfied.  As for the 2DEG in quantum wells, we find that the DP-BIA mechanism generally dominates the spin relaxation, except in those special cases in which it is absent for symmetry reasons. 

We now estimate some spin relaxation times for GaAs in specific situations. For a 3D donor concentration of 
$n_{3\mathrm D}=1.5\times 10^{17}$~$\mathrm{cm}^{-3}$ we find (see Fig.~3-23 in Ref. \cite{Streetman}) that the corresponding mobility is $\mu =4\times 10^3\mathrm{cm}^2/(\mathrm{Vs})$,
which corresponds to a 3D elastic relaxation time~$\tau =\mu m^*/e=1.5\times 10^{-13}$s.
If all these donors were interstitials, with $l_1$ given above in Table~\ref{estab}, then for a quantum well with a Fermi energy  of $34$me\!V, the 
spin relaxation time would be $9\times 10^{-11}$s.
In bulk GaAs  at $E_{\mathrm F}=26$me\!V we estimate the spin relaxation time from this mechanism under these same assumptions as $2\times 10^{-10}$s.
\newcolumntype{A}{>{$}m{10ex}<{$}}
\newcolumntype{B}{>{$}m{4.5ex}<{$}}
  \begin{table}
    \begin{tabular}{%
|A      |  B  |     B|    B|     B|     B|  B  |   }\hline
        &\!\!\!\text{GaAs} &\!\!\!\text{GaSb}  &\text{InP}  &\!\!\text{InAs}  &\!\!\text{ZnSe}  &\!\!\!\text{CdTe}\tabularnewline\hline
E_g(\mathrm{eV}) &1.52 &0.73  &1.34 &0.35  &2.82  &1.5 \tabularnewline\hline
\Delta (\mathrm{eV})   &0.34 &0.80  &0.11 &0.41  &0.43  &0.85\tabularnewline\hline
P(\mathrm{eV}\cdot \AA{}) &10.5 &9.69  &8.45 &9.01  & 9.6  &8.9\tabularnewline\hline
l_1(\AA{})   &0.84 &4.7   &0.32& 9.1    & 0.3  &1.4 \tabularnewline\hline
l_2(\AA{})   &3.55 &9.6  &1.96 &18    & 1.38 &3.87 \tabularnewline\hline
b(\mathrm{eV}\cdot \AA{}^3)&27.6& 108 & 4.86& 34.3& 13.6& 41.8 \tabularnewline\hline
C    &0.1& 0.5& 0.06&\!\! 0.004& 0.4& 0.16  \tabularnewline\hline
    \end{tabular}
\caption{Properties of several  direct-gap zinc-blende semiconductors.
Here we denote $b\equiv b_{41}^{6c6c}$ -- the spin-orbit parameter defined in~\cite{Winkler}.
The constant~$C$ is defined after~\eqref{vle}.
\label{estab}}\end{table}

\section{Conclusion}
The reduction of the local point-group symmetry of an atomic cell caused, e.g., by interstitial atoms,
leads to a new type of spin-orbit interaction~[Eq.~\eqref{novelSOI}]  which scales inversely to gap size in direct-gap zinc-blende (A$_3$B$_5$) semiconductors.
We speculate that low-symmetry defects may also play a role in enhancing the spin-orbit interaction in group IV semiconductors, and at polar interfaces between materials of cubic symmetry, such as at the LaAlO$_3$/SrTiO$_3$ interface~\cite{Caviglia2010}.  Other quantities, which originate from off-diagonal  matrix elements of the $\vec k\cdot \vec p$   Hamiltonian (\emph{e.g.}, the renormalization of the electron mass and of the $g$ factor) will also be affected. 

\acknowledgments
We acknowledge the support of ARO MURI through Grant No. W911NF-08-1-0317.
\bibliographystyle{eplbib}
\bibliography{main}

\begin{thebibliography}{10}
\expandafter\ifx\csname url\endcsname\relax\def\url#1{\texttt{#1}}\fi

\bibitem{Blount1962}
\Name{Blount E.~I.} \REVIEW{Solid State Physics}{13}{1962}{305}.

\bibitem{Winkler}
\Name{Winkler R.} \Book{Spin-Orbit Effects in Two-Dimensional Electron and Hole
  Systems} (Springer) 2003.

\bibitem{Engel}
\Name{Engel H.-A., Halperin B.~I. \and Rashba E.~I.} \REVIEW{Phys. Rev.
  Lett.}{95}{2005}{166605}.

\bibitem{Awschalom2002}
\Name{Awschalom D.~D., Samarth N. \and Loss D.} (Editors) \Book{Semiconductor
  Spintronics and Quantum Computation} (Springer Verlag, Heidelberg) 2002.

\bibitem{Awschalom2007}
\Name{Awschalom D.~D. \and Flatt\'e M.~E.} \REVIEW{Nature
  Physics}{3}{2007}{153}.

\bibitem{PikusTitkov}
\Name{Pikus G.~E. \and Titkov A.~N.} \Book{Spin relaxation of charge cariers
  with optical orientation in semiconductors} in \Book{Optical orientation},
  edited by \Name{Zaharchenya B.~P. \and Meier F.} (Nauka (Leningrad)) 1989
  Ch.~3 pp. 62--108.

\bibitem{YuCardona}
\Name{Yu P.~Y. \and Cardona M.} \Book{Fundamentals of semiconductors} 3rd
  Edition (Springer) 2001.

\bibitem{Elliott1954}
\Name{Elliott R.~J.} \REVIEW{Phys. Rev.}{96}{1954}{266}.

\bibitem{Yafet1963}
\Name{Yafet Y.} \REVIEW{Solid State Physics}{14}{1963}{2}.

\bibitem{Lau2001}
\Name{Lau W.~H., Olesberg J.~T. \and Flatt\'e M.~E.} \REVIEW{Phys. Rev.
  B}{64}{2001}{161301(R)}.

\bibitem{PhysRevB.64.201301}
\Name{Olesberg J.~T., Lau W.~H., Flatt\'e M.~E., Yu C., Altunkaya E., Shaw
  E.~M., Hasenberg T.~C. \and Boggess T.~F.} \REVIEW{Phys. Rev.
  B}{64}{2001}{201301}.

\bibitem{Wu201061}
\Name{Wu M., Jiang J. \and Weng M.} \REVIEW{Physics Reports}{493}{2010}{61 }.

\bibitem{ActaFabian}
\Name{Fabian J., Matos-Abiague A., Ertler C., Stano P. \and Zutic I.}
  \REVIEW{Acta Physica Slovaca}{57}{2007}{565}.

\bibitem{Dyakonov1986}
\Name{D'yakonov M.~I. \and Kachorovskii V.~Y.} \REVIEW{Soviet Physics
  Semiconductors}{20}{1986}{110}.

\bibitem{Pantelides-dc}
\Name{Pantelides S.~T.} (Editor) \Book{Deep centers in semiconductors: a state
  of the art approach} 2nd Edition (Gordon and Breach, Philadelphia) 1992.

\bibitem{Queisser14081998}
\Name{Queisser H.~J. \and Haller E.~E.} \REVIEW{Science}{281}{1998}{945}.

\bibitem{montambauxBook}
\Name{Akkermans E. \and Montambaux G.} \Book{Mesoscopic Physics of Electrons
  and Phonons} (Cambridge University Press, New York) 2007.

\bibitem{approxes}
We have checked that the off-diagonal elements of~${\hat{\bar H}}_v$ give
  negligible contributions (central-cell corrections) to both Rashba SOI and
  the novel one (if~$Δ≠0$). We do not provide here the details of this
  check, which allows us to ignore the contribution of the off-diagonal
  elements of~${\hat{\bar H}}_v$.

\bibitem{allConsidered}
In other words, expression \eqref{sumRule} is spin-independent when ${\hat{\bar
  H}}_{v8}={\hat{\bar H}}_{v7}$ \emph{within the validity} of the~$8×8$ $\vec
  k·\vec p$ model.

\bibitem{ahiezerEng}
\Name{Akhiezer A.~I. \and Peletminsky S.~V.} \Book{Methods of Statistical
  Physics} (Pergamon Press, Oxford) 1981.

\bibitem{bigCond}
\Name{Chalaev O. \and Loss D.} \REVIEW{Phys. Rev. B}{80}{2009}{035305}.

\bibitem{Streetman}
\Name{Streetman B.~G. \and Banerjee S.} \Book{Solid state electronic devices}
  6th Edition (Prentice Hall) 2006.

\bibitem{Caviglia2010}
\Name{Caviglia A.~D., Gabay M., Gariglio S., Reyren N., Cancellieri C. \and
  Triscone J.-M.} \REVIEW{Phys. Rev. Lett.}{104}{2010}{126803}.

\end{thebibliography}
\end{document}